\documentclass[submission, Phys]{SciPost}
\usepackage{bm,bbold}

\newcommand{\tr}{{\rm tr}}

\newcommand{\be}{\begin{equation}}
\newcommand{\ee}{\end{equation}}

\begin{document}

\begin{center}{\Large \textbf{
Closed hierarchy of Heisenberg equations\\
in integrable models with Onsager algebra
%in the quantum transverse-field Ising chain
}}\end{center}

% TODO: write the author list here. Use initials + surname format.
% Separate subsequent authors by a comma, omit comma at the end of the list.
% Mark the corresponding author with a superscript *.
\begin{center}
Oleg Lychkovskiy\textsuperscript{1,2,3},
\end{center}

\begin{center}
{\bf 1} Skolkovo Institute of Science and Technology, \\ Bolshoy Boulevard 30, bld. 1, Moscow 121205, Russia
\\
{\bf 2} Laboratory for the Physics of Complex Quantum Systems, \\ Moscow Institute of Physics and Technology,\\ Institutsky per. 9, Dolgoprudny, Moscow  region,  141700, Russia
\\
{\bf 3} Department of Mathematical Methods for Quantum Technologies,\\
		Steklov Mathematical Institute of Russian Academy of Sciences\\
		8 Gubkina St., Moscow 119991, Russia
\\
* o.lychkovskiy@skoltech.ru
\end{center}

\begin{center}
\today
\end{center}

% For convenience during refereeing: line numbers
%\linenumbers

\section*{Abstract}
{\bf
Dynamics of a quantum system can be described by coupled Heisenberg equations. In a generic many-body system these equations form an exponentially large hierarchy that is intractable without approximations. In contrast, in an integrable system a small subset of operators can be closed with respect to commutation with the Hamiltonian. As a result, the Heisenberg equations for these operators can form a smaller closed system amenable to an analytical treatment. We demonstrate that this  indeed happens in a class of integrable models where the Hamiltonian is an element of the Onsager algebra. We explicitly solve the system of Heisenberg equations for operators from this algebra. Two specific models are considered as examples: the transverse field Ising model and the superintegrable chiral 3-state Potts model.
%  This allows us to describe the dynamics of corresponding observables for an arbitrary initial state  This allows one to describe the dynamics of the corresponding observables for an arbitrary initial state. As a particular example, we calculate time-dependent expectation values $\langle \sigma^z_j \rangle_t$, $\langle \sigma^x_j \sigma^x_{j+1} \rangle_t$, $\langle \sigma^y_j \sigma^y_{j+1} \rangle_t$ and  $\langle \sigma^x_j \sigma^y_{j+1} \rangle_t$
%(where the field is directed along the $z$-axis)
%for a translationally  invariant product initial state with an arbitrarily directed polarization.
}

\vspace{10pt}
\noindent\rule{\textwidth}{1pt}
\tableofcontents\thispagestyle{fancy}
\noindent\rule{\textwidth}{1pt}
\vspace{10pt}

\section{Introduction}
\label{sec:intro}
% TODO: write your article here.

Describing an out-of-equilibrium dynamics of a quantum many-body system is, in general, a formidable task. One can expect that this task is considerably simplified for integrable models. However, the integrability turns out to be somewhat deceptive when it comes to the dynamics. A range of methods for addressing the dynamics and resulting non-thermal steady states is under active development, including the quench action approach \cite{Caux_2013_Quench_action,Caux_2016}, generalized Gibbs ensemble \cite{Rigol_2007_GGE},  generalized hydrodynamics \cite{Castro-Alvaredo_2016_GHD,Bertini_2016_GHD} and advanced techniques for summing  form-factor expansions \cite{Burovski2013,Gamayun_2018_Impact,Granet_2020_Finite}. Nevertheless, each instance of an analytical calculation in this field is a certain tour de force.

In the present paper we address the out-of-equilibrium many-body dynamics via a system of coupled Heisenberg equations. In a generic many-body model, such system of equations consists of exponentially many equations involving progressively nonlocal operators, analogously to the closely related Bogoliubov–Born–Green–Kirkwood–Yvon hierarchy of equations on reduced density matrices \cite{Bogoliubov_1946,Bogoliubov_1947,Born_1947,Kirkwood_1946,Yvon_1957}. However, one can expect that in an integrable system a subset of operators can turn out to be closed with respect to the commutation with the Hamiltonian, and, as a result, the Heisenberg equations for these operators decouple from the rest of the equations. This indeed happens for a quadratic fermionic (or bosonic) Hamiltonian, where for any fixed integer $k$ a set of operators spanned by products of $k$ fermionic (or bosonic) creation and annihilation operators  remains closed with respect to commutation with the Hamiltonian. This property of quadratic Hamiltonians can also be used to describe the dynamics  spin-$1/2$ models that can be mapped to systems of noninteracting fermions. For example, in Ref. \cite{brandt1976exact} an open-end transverse field Ising model was considered, and a set of Heisenberg equations for operators linear in terms of fermionic operators ($k=1$) was derived and solved.

%These operators turned out to be either highly nonlocal (in terms of spin observables) or resided close to the boundary. To asses the bulk properties of the model a product of two operatore.

Here we study models with the Hamiltonian of the form
\be\label{H general}
H=a_0 \, A^0+ a_1 \, A^1,
\ee
where $a_0$, $a_1$ are real numbers and  $A_0$, $A_1$ are two operators that satisfy the Dolan-Grady conditions \cite{Dolan_Grady_1982}
\begin{align}\label{Dolan-Grady}
  [A_0,[A_0,[A_0,A_1]]] & = 16 [A_0,A_1], \nonumber \\
  [A_1,[A_1,[A_1,A_0]]] & = 16 [A_1,A_0].
\end{align}
This pair of operators generates the Onsager algebra of operators \cite{Onsager_1944} (as briefly reviewed below in Sec. \ref{subsec: algebra}).  This class of models includes the transverse-field Ising model \cite{nambu1950note,Lieb_1961,Katsura_1962,Pfeuty_1970} and superintegrable chiral Potts models \cite{vonGehlen_1985}. The latter are not reduced to noninteracting fermions.

We derive and solve a system of Heisenberg equations for a set of operators $G^n$, $A^n$ spanning the Onsager algebra. This is done in the next section. Then, in Sec. \ref{sec: Ising}, we apply the  general solution to describe the out-of-equilibrium dynamics of the transverse field Ising model for a translation-invariant product initial state polarized in an arbitrary direction. In Sec. \eqref{sec: Potts} we address the out-of-equilibrium dynamics of the 3-state Potts model. We conclude by the summary and outlook in Sec. \ref{sec: discussion}.

\section{Heisenberg equations and their solution \label{sec: Onsager}}

\subsection{Onsager algebra \label{subsec: algebra}}

The Onsager algebra \cite{Onsager_1944} is a Lie algebra spanned by operators  $G^n$, $A^n$ that are recursively generated starting from $A^0$, $A^1$ according to
\begin{align}\label{generating algebra}
   G^n&= \frac14 [A^n,A^0], & n=0,1,2,\dots,\nonumber \\
%  G^1 & =\frac14 [A^1,A^0],  &\\
  A^{n+1} -A^{n-1} &=  \frac12 [G^1, A^n], & n=0,\pm1,\pm2,\dots
%  A^{m-1} &= A^{m+1} - \frac12 [G_1, A_m], & m=0,-1,-2,\dots \\
\end{align}
Commutation plays the role of the bilinear product of the algebra, with the structure explicitly given by
\begin{align}\label{structure of algebra}
[A^n, A^m] & =4\, G^{n-m}, \nonumber \\
[G^n, A^m] & =2 A^{m+n} - 2 A^{m-n}, \nonumber \\
[G^n, G^m] & =0
\end{align}
with  $G^{-n}=-G^n$. The Dolan-Grady conditions \eqref{Dolan-Grady} are necessary and sufficient for the set of operators $A^n$, $G^n$ generated by the recursion  \eqref{generating algebra} to satisfy  eq.~\eqref{structure of algebra} \cite{Dolan_Grady_1982,perk1989star,Davis_1991}.

Originally, the Onsager algebra emerged in studies of the classical Ising model \cite{Onsager_1944}. An apparently first explicit construction of an Onsager algebra representation for a quantum model -- namely, for the $XY$ spin chain -- was presented in \cite{Jha_1973}. In an important paper by von Gehlen and Rittenberg \cite{vonGehlen_1985} it was shown that for an arbitrary positive integer~$n$ a nearest-neighbour spin-$n/2$ Hamiltonian of the form \eqref{H general} can be constructed, with $A^0$ and $A^1$ satisfying the Dolan-Grady conditions \eqref{Dolan-Grady} and thus generating a representation of the Onsager algebra. The spin-$1/2$ case is the transverse-field Ising model. The higher spin models are known as $n$-state  superintegrable chiral Potts models or $Z_n$-invariant clock models. This discovery triggered a considerable lasting interest in such models \cite{Albertini_1989,Davies_1990,nishino2008algebraic,Vernier_2019} and, more generally, in applications of the Onsager algebra and its generalizations to  quantum integrability \cite{uglov1996sl,Baseilhac_2005,Baseilhac_2013_half-infinite,Gritsev_2017,baseilhac2018frt}.

\subsection{Solving Heisenberg equations}

We will work in the Heisenberg representation where a time-dependent expectation value of an observable $\cal O$ is given by
\be\label{expectation value}
\langle {\cal O} \rangle_t = \tr\, \rho_0 \, {\cal O}_t,
\ee
with $\rho_0$ being the initial state of the system and ${\cal O}_t$ -- the Heisenberg operator related to the Schr\"odinger operator ${\cal O}$ as
\be\label{Heisenberg picture}
{\cal O}_t = e^{i H t} {\cal O} e^{- i H t}.
\ee
The Heisenberg operator satisfies the Heisenberg equation of motion,
\be\label{Heisenberg equation}
\partial_t {\cal O}_t = i[H,{\cal O}_t],
\ee
with the initial condition ${\cal O}_{0}={\cal O}$.

%Our strategy is to solve coupled Heisenberg equations for Heisenberg string operators  $X^n_t$, $Y^n_t$ and $M^n_t$ and then to calculate their time-dependent expectation values. The present subsection is devoted to the former task, while the next subsection - to the latter one.

We are going to solve Heisenberg equations for Heisenberg operators $G^n_t$, $A^n_t$. These equations are straightforwardly obtained from eqs.  \eqref{H general}, \eqref{structure of algebra}:

\begin{align}\label{system of equations}
\partial_t G^n_t & = 2 \,i\, \Big( a_0(-A^n_t+A^{-n}_t)+a_1(-A^{1+n}_t+A^{1-n}_t)\Big),\nonumber \\[1.5ex]
\partial_t A^n_t & =-4 \, i \, \Big( a_0\, G^n_t+ a_1 \, G^{n-1}_t \Big),
\end{align}
where $n=0,\pm1,\pm2,\dots$

We exclude $A^n_t$ from eq. \eqref{system of equations} and obtain a system of the second-order equations:
\begin{align}\label{system of equations 2}
\partial^2_t G^n_t & = -16\Big(a_0 \, a_1 \, G^{n-1}_t+(a_0^2 + a_1^2)G^n_t+a_0 \, a_1\, G^{n+1}_t\Big),\qquad n =1,2,\dots
\end{align}
This system can be conveniently written in the compact matrix notations as
\be\label{ODE matrix form}
\partial^2_t G_t =T G_t,
\ee
where $G_t$ is a vector composed of $G^n_t$ and $T$ is a   tridiagonal Toeplitz matrix with matrix elements $T_{nn}=- 16(a_0^2 + a_1^2)$ and $T_{n \,n\pm1}=-16\,a_0 \, a_1$.

To proceed further, we truncate the matrix $T$ to a finite $(N-1)\times (N-1)$ matrix, keeping the same notation for the truncated matrix. In fact, this truncation emerges naturally for finite-dimensional representations of Onsager algebra \cite{Davies_1990}, where $A^n, G^n$ are operators on the Hilbert space of a spin chain with $N$ spins, as discussed in more detail in the next section. Anyway, the dependence on the system size goes away for local observables in the thermodynamic limit of $N\rightarrow\infty$.

 We then diagonalize the truncated matrix by a unitary transformation $U$,
\be\label{diagonalization}
T=-U\, {\cal E}^2 \, U^\dagger,
\ee
where the matrix elements of $U$ read
\be\label{Umn}
U_{mn}=\sqrt{\frac2{N}} \, \sin\frac{\pi mn}N,\qquad m,n=1,\dots,(N-1),
\ee
and ${\cal E}$ is a diagonal matrix with diagonal entries
\be\label{varepsilon}
\varepsilon_{\varphi_n}=4\sqrt{a_0^2 + a_1^2+2a_0 \, a_1 \cos\varphi_n},\quad \varphi_n=\frac{\pi n}N, \qquad n=1,2,\dots,(N-1).
\ee
%Remarkably, $\varepsilon_{\varphi_n}$ is twice the energy of  elementary fermionic excitations of the model \cite{Pfeuty_1970,Essler_2016}.

By a standard argument, the diagonalization \eqref{diagonalization} implies the following solution of eq.~\eqref{system of equations 2}:
\be\label{solution}
G^n_t=\sum_{m=1}^{\infty} \Big( \partial_t c^{nm}_t \, G^m+ 2\,i\,c^{nm}_t \big(a_0(-A^m+A^{-m})+a_1(-A^{1+m}+A^{1-m})\big)\Big)
\ee
with
\be\label{c_t}
%c^{nm}_t=\frac2N \sum_{l=1}^{N-1}  \sin (n\varphi_l) \sin (m\varphi_l)\frac{\sin (\varepsilon_{\varphi_l} t)}{\varepsilon_{\varphi_l}}.
c^{nm}_t\equiv\frac2\pi \int_0^\pi d\varphi  \sin (n\varphi) \sin (m\varphi)\frac{\sin (\varepsilon_{\varphi} t)}{\varepsilon_{\varphi}},
\ee
where the limit $N\rightarrow \infty$ is already taken. $A^n_t$ is then found from the second line of eq. \eqref{system of equations} and reads
\begin{align}\label{solution An}
A^n_t= & A^n-4\,i\,\sum_{m=1}^{\infty} \Bigg( \big(a_0 \,c^{nm}_t+a_1 \, c^{(n-1)\; m}_t\big) \, G^m \nonumber \\
 & + 2 \,i\,\big(a_0 \,C^{nm}_t+a_1 \, C^{(n-1)\; m}_t\big) \Big( a_0(-A^m+A^{-m})+a_1(-A^{1+m}+A^{1-m})\Big)\Bigg),
\end{align}
where
\be\label{C_t}
C^{nm}_t \equiv\int_0^t dt' c^{nm}_{t'}=\frac2\pi \int_0^\pi d\varphi  \sin (n\varphi) \sin (m\varphi)\frac{1-\cos (\varepsilon_{\varphi} t)}{\varepsilon_{\varphi}^2}.
\ee

Eqs. \eqref{solution}-\eqref{C_t} are the main general results of the present paper. They express time-dependent Heisenberg operators $G^n_t$, $A^n_t$ through the Schrodinger operators $G^n$, $A^n$.

%\subsection{Heisenberg operators at equilibrium}

%One can easily take the limit $t\rightarrow \infty$ in eqs. \eqref{solution}, \eqref{solution An} to obtain
%\begin{align}\label{solution An}
%G^n_\infty=& \, 0, \nonumber \\
%A^n_\infty= & \,  A^n+8\,\sum_{m=1}^{\infty} \big(a_0 \,C^{nm}_\infty+a_1 \, C^{(n-1)\; m}_\infty \big) \Big( %%a_0(-A^n+A^{-n})+a_1(-A^{1+n}+A^{1-n})\Big)\Bigg),
%\end{align}
%where
%\be\label{Cinfty}
%C^{nm}_\infty \equiv\int_0^t dt' c^{nm}_{t'}=\frac2\pi \int_0^\pi d\varphi  \frac{\sin (n\varphi) \sin (m\varphi)}{\omega_{\varphi}^2}.
%\ee

%This solution is valid for any finite $N$ as well as in the thermodynamic limit $N\rightarrow \infty$. In the latter case the sum in eq. \eqref{c_t} should be replaced by the integral. The solutions for $X^n_t$ and  $Y^n_t$ are easily obtained from eq. \eqref{solution} by integrating the first two lines of eq. \eqref{system of equations}.

%\subsection{Time-dependent Hamiltonian}

%\be\label{Htime-dependent}
%H=a_0(t) A^0+ a_1(t) A^1,
%\ee

\section{Out-of-equilibrium dynamics of the Ising model \label{sec: Ising}}

In the present section we apply our general results to the transverse field Ising model. This model belongs to the simplest class of {\it noninteracting} integrable spin chains since it can be mapped to noninteracting fermions \cite{nambu1950note,Lieb_1961,Katsura_1962,Pfeuty_1970}. Early studies of the transverse field Ising model mostly addressed equilibrium instantaneous  and time-dependent correlation functions \cite{Pfeuty_1970,Niemeijer_1967,Katsura_1970,Mazur_1973,Lajzerowicz_1975,Tommet_1975,brandt1976exact,Capel_1977_autocorrelation,Vaidya_1978} (this topic still attracts a deal of attention \cite{Gohmann_2020_High-temperature,Granet_2020_Finite,gamayun2020effective}). The studies of the  out-of-equilibrium dynamics, while initiated in notable early papers \cite{Niemeijer_1967,Barouch_1970}, came into focus much later \cite{Igloi_2000_long-range,Amico_2004,Sengupta_2004_quench,Chang_2010_Dynamics,Calabrese_2011,Calabrese_2012,Fagotti_2013_reduced,Granet_2020_Finite}, with most efforts directed to the dynamics after a quantum quench, as reviewed e.g. in \cite{Essler_2016}. Here we describe the out-of-equilibrium dynamics for a previously unstudied initial condition specified below.

%The Hamiltonian of the translation-invariant transverse field Ising model is given by  with the Hamiltonian

We consider the transverse field Ising chain with $N$ spin $1/2$. The Hamiltonian is given by eq. \eqref{H general}, where the finite-dimensional representation of the Onsager algebra reads (see e.g.  \cite{perk2017onsager})
\begin{align}\label{string operators}
G^n= & \frac{i}{2} \sum_{j=1}^N \left( \sigma^x_j \left(\prod_{m=1}^{n-1} \sigma^z_{j+m} \right) \sigma^y_{j+n} + \sigma^y_j \left(\prod_{m=1}^{n-1} \sigma^z_{j+m} \right) \sigma^x_{j+n} \right), \nonumber \\[1.5ex]
A^n= &  \sum_{j=1}^N \sigma^x_j \left(\prod_{m=1}^{n-1} \sigma^z_{j+m} \right) \sigma^x_{j+n}, \nonumber \\[1.5ex]
A^{-n}= & \sum_{j=1}^N \sigma^y_j \left(\prod_{m=1}^{n-1} \sigma^z_{j+m} \right) \sigma^y_{j+n}, \nonumber \\[1.5ex]
  G^{0}= &\, G^{N}= 0, \qquad A^0= - \sum_{j=1}^N \sigma^z_j, \qquad  A^{N}=  \sum_{j=1}^N \left(\prod_{m\neq j} \sigma^z_{m}\right),  \nonumber \\[1.5ex]
G^{-n}= &\, G^n, \qquad A^{n\pm 2N}=A^n,\qquad G^{n\pm 2N}=G^n.
\end{align}
Here $\sigma_j^{x,y,z}$ are Pauli matrices, $n=1,2,\dots,(N-1)$,  all operators $G^n$, $A^n$ (as well as the Hamiltonian) are assumed translation invariant with $\sigma^{\alpha}_{N+n}\equiv\sigma^{\alpha}_n$, and $\prod_{m=1}^{n-1} \sigma^z_{j+m}$ is void (i.e. replaced by $1$) for $n=1$.

Note that, thanks to the equality $G^N=0$, the first $(N-1)$ equations in  eq. \eqref{system of equations 2} decouple from the subsequent equations, which rigorously justifies the truncation of the matrix $T$ in  eq. \eqref{ODE matrix form}. This also implies that our results, though presented here in the thermodynamic limit, are also valid, upon a straightforward modification, for any finite~$N$.

Let us remark that the ``string'' operators \eqref{string operators}  recurrently show up in the studies of the Ising model \cite{Pfeuty_1970} and related models, the topics varying from integrable Floquet dynamics \cite{Prozen_2000_exact} to a variational ansatz for nonintegrable spin chains  \cite{Wurtz_2020_emergent} to adiabatic gauge potentials  \cite{hatomura2020controlling}.

\begin{figure}[t] %  figure placement: here, top, bottom, or page
		\centering
\includegraphics[width=\linewidth]{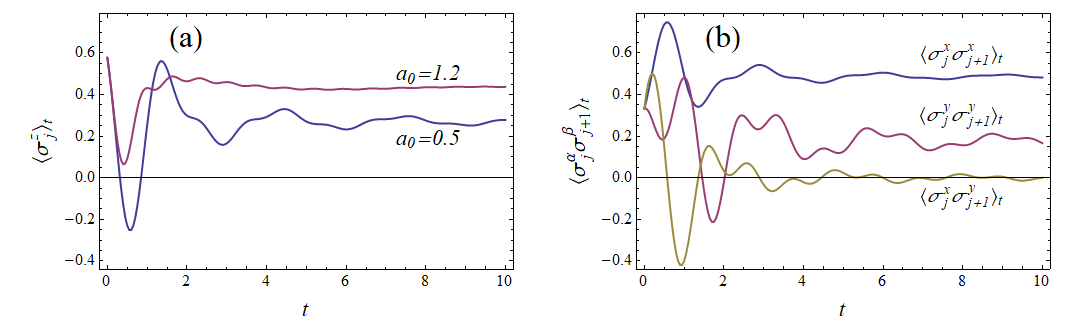}
        \caption{Time dependence of observables in the transverse field Ising model, see eqs.~\eqref{observables}-\eqref{observables for fig}. The initial state is a  product translation-invariant  state \eqref{initial condition} with the  polarization $p_x=p_y=p_z=1/\sqrt3$. (a)  $\langle \sigma^z_j \rangle_t$ for $a_1=-1$ and two values of $a_0$ specified in the plot.  (b) $\langle \sigma^x_j \sigma^x_{j+1} \rangle_t$, $\langle \sigma^y_j \sigma^y_{j+1} \rangle_t$ and $\langle \sigma^x_j \sigma^y_{j+1} \rangle_t$ for $a_1=-1$ and $a_0=0.5$. }
		\label{fig}
\end{figure}

We consider the dynamics of the Ising model prepared in a product initial state
\be\label{initial condition}
\rho_0=\prod_{m=1}^{N} \left( \frac12(1+{\bf p}{\bm \sigma})\right),
\ee
where ${\bf p}=(p_x,p_y,p_z)$ is the polarization vector inside the Bloch sphere,  ${\bf p}^2 \leq 1$.  While for the initial polarisation along the $z$-axis the problem has been studied earlier \cite{Niemeijer_1967}, for a generic polarisation the dynamics has remained unexplored.
%This initial condition has recently attracted attention since it gives rise to a periodic trajectory with a short period in a classical translation-invariant spin system \cite{Ermakov_2020}.

We wish to calculate time-dependent expectation values $\langle G^n\rangle_t$, $\langle A^n\rangle_t$. To this end we plug the solution \eqref{solution}--\eqref{C_t} of the Heisenberg equations to eq. \eqref{expectation value}.  The initial condition \eqref{initial condition} implies
\be
\langle A^n \rangle_0=N p_x^2\, p_z^{n-1},\quad \langle A^{-n} \rangle_0 =N p_y^2\, p_z^{n-1},\quad \langle G^n \rangle_0 =i N p_x \, p_y\, p_z^{n-1},\quad n=1,2,\dots
\ee
and $\langle A^0 \rangle_0 = -N p_z$, $\langle A^N \rangle_0 = N p_z^{N-1}$. We plug these values into eq. \eqref{solution} and explicitly take the sum over $m$, which is essentially a geometric series. This way we obtain in the thermodynamic limit
\begin{align}
\langle A^n\rangle_t = & \langle A^n \rangle_0 +4 N  \int_0^\pi \frac{d\varphi}{\pi} \Big(a_0 \sin\big(n\varphi\big)+ a_1 \sin\big( (n-1) \varphi\big)\Big)\, \sin \varphi
\left(
R_\varphi \frac{\sin\varepsilon_\varphi t}{\varepsilon_\varphi} + Q_\varphi \frac{1-\cos\varepsilon_\varphi t}{\varepsilon_\varphi^2}
\right),\nonumber \\[1.5ex]
\langle G^n\rangle_t  = & i N \int_0^\pi \frac{d\varphi}{\pi} \sin (n\varphi)\, \sin \varphi
\left(
R_\varphi \cos\varepsilon_\varphi t + Q_\varphi \frac{\sin\varepsilon_\varphi t}{\varepsilon_\varphi}
\right),\label{observables}
\end{align}
where
\begin{align}
R_\varphi & =\frac{2 \, p_x \, p_y}{1+p_z^2-2p_z \,\cos \varphi}, \nonumber \\
Q_\varphi & =-4\,a_1 \left( \frac{p_x^2\,p_z-p_y^2/p_z+(a_0/a_1)(p_x^2-p_y^2)}{1+p_z^2-2p_z \cos \varphi}+ p_y^2/p_z+p_z \right).
\end{align}
In Fig. \ref{fig} we plot the evolution of
\begin{equation}\label{observables for fig}
\begin{array}{l}
\langle \sigma^z_j \rangle_t = - N^{-1} \langle A^0\rangle_t,  \\[1.5ex]
\langle \sigma^x_j \sigma^x_{j+1} \rangle_t = N^{-1} \langle A^1\rangle_t,\quad
\langle \sigma^y_j \sigma^y_{j+1} \rangle_t = N^{-1} \langle A^{-1}\rangle_t,\quad
\langle \sigma^x_j \sigma^y_{j+1} \rangle_t = -i N^{-1} \langle G^1\rangle_t.
\end{array}
\end{equation}

We have verified that in a special case of $p_z=\pm 1$, $p_x=p_y=0$ our result for  $\langle \sigma^z_j \rangle_t$ coincides with that of Niemeijer \cite{Niemeijer_1967}. Further, very recently the dynamics starting from the state with  $p_x=1$, $p_y=p_z=0$ has been studied in detail in ref. \cite{wu2021exact}. An analytical formula  for $\langle \sigma^x_j \sigma^x_{j+1} \rangle_t$ provided in  \cite{wu2021exact} agrees with our result.

%Remarkably, $\varepsilon_{\varphi_n}$ is twice the energy of  elementary fermionic excitations of the model \cite{Pfeuty_1970,Essler_2016}.

Let us emphasise that the results \eqref{observables} have been obtained without resorting to the fermionic picture of the Ising model. Undoubtedly, the same results could be obtained in the fermionic language, since the string  operators \eqref{string operators} are just quadratic operators in terms of fermions \cite{Jha_1973}. However, our method is  more straightforward, both conceptually and technically. Furthermore, it avoids known difficulties of the fermionic approach related to the boundary term dependent on the overall parity $\prod_j \sigma_j^z $\cite{Capel_1977_autocorrelation}, as well as to dealing with initial states not amenable to the generalized Wick's theorem \cite{Essler_2016}.

%The Heisenberg equations were previously derived and solved for a different set of string operators in the open-end Ising chain \cite{brandt1976exact}. These operators  are linear in terms of fermions. In contrast to the operators considered here, the set of operators in  \cite{brandt1976exact} does not contain local (in terms of spins) operators in the bulk.

We also note that for equilibrium time-dependent correlation functions a set of {\it nonlinear} equations has been derived \cite{Perk_1980,Perk_1984} and solved numerically \cite{perk2009new}. This method relies on the Wick's theorem and, therefore, can not be generalized to the dynamics starting from non-Gaussian initial states.

\section{Out-of-equilibrium dynamics of the 3-state Potts model \label{sec: Potts}}

\begin{figure}[t] %  figure placement: here, top, bottom, or page
		\centering
\includegraphics[width=\linewidth]{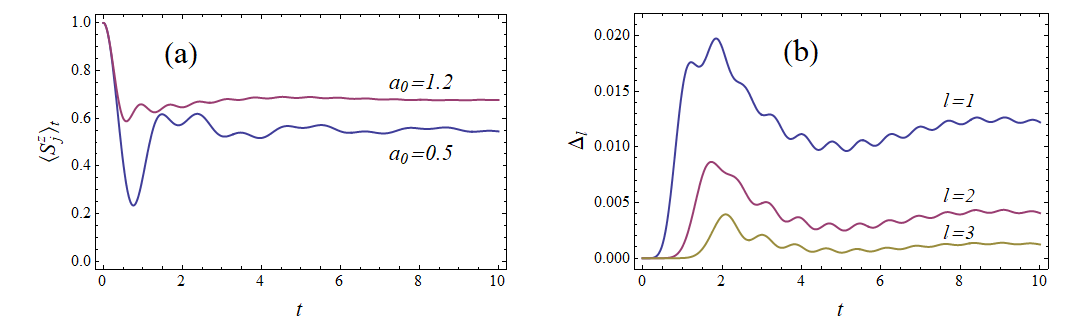}
        \caption{(a) $\langle S_j^z \rangle_t$ in the superintegrable chiral 3-state Potts model calculated via eq. \eqref{Sz Potts} with the sum truncated at $l_{\rm max}=4$. The initial state is the  product translation-invariant  state~\eqref{initial condition Potts} with all spins polarized along the $z$-direction. The calculation is done for  $a_1=-1$ and two values of $a_0$ shown in the plot. (b) Illustration of the convergence of the series in eq. \eqref{Sz Potts}. Plotted is the difference $\Delta_l$ between two approximations to $\langle S_j^z \rangle_t$ derived from the truncated eq.~\eqref{Sz Potts}, with truncation orders equal to  $l_{\rm max}=l$ and $l_{\rm max}=4$, respectively. The parameters of the Hamiltonian are $a_0=1.2$ and $a_1=-1$. One can see that truncation error is small already for $l_{\rm max}=1$ and rapidly vanishes with growing $l_{\rm max}$.}
		\label{fig: Potts}
\end{figure}

Here we consider the  superintegrable  chiral 3-state Potts model \cite{vonGehlen_1985}. It is a model of $N$ spins-$1$ with the Hamiltonian \eqref{H general}, where
\begin{align}\label{A0 A1 Potts}
A^0=&\,\frac43\sum_{j=1}^N \left(\frac{\tau_j}{1-\omega^*}+h.c.\right) \nonumber\\
A^1=&\,\frac43\sum_{j=1}^N \left(\frac{\sigma_j \sigma_{j+1}^\dagger}{1-\omega^*}+h.c.\right).
\end{align}
Here $\omega=e^{2\pi i/3}$ and $\tau_j$, $\sigma_j$ are operators acting on the $j$'th spin with the following properties:
\be\label{commutations relations}
\tau_j^3=\mathbb{1}_j,\qquad \sigma_j^3=\mathbb{1}_j,\qquad \tau_j^2=\tau_j^\dagger,\qquad \sigma_j^2=\sigma_j^\dagger,\qquad \sigma_j \tau_j=\omega \, \tau_j \sigma_j.
\ee
Analogously to the previous section, the system is assumed to be translation invariant with $\tau_{j+N}\equiv\tau_j$,  $\sigma_{j+N}\equiv\sigma_j$. This model, in contrast to the Ising spin-$1/2$ chain considered above, does not map to noninteracting fermions.

We choose the following matrix representation of these operators (see e.g. \cite{Vernier_2019}):
\be\label{tau sigma}
\tau=
\begin{pmatrix}
  1 & 0 & 0 \\
  0 & \omega & 0 \\
  0 & 0 & \omega^*
\end{pmatrix},
\qquad
\sigma=
\begin{pmatrix}
  0 & 1 & 0 \\
  0 & 0& 1 \\
  1 & 0 & 0
\end{pmatrix}.
\ee
With this choice, the $z$-projection of a spin, $S^z\equiv{\rm diag}(1,0,-1)$, can be expressed as $S^z=((1-\omega^*)^{-1} \, \tau+h.c.)$, and therefore $A^0$ is, up to the factor $4/3$, a total polarization along the $z$-axis:
\begin{align}\label{A0 Potts}
A^0=&\,\frac43\sum_{j=1}^N S^z_j.
\end{align}

Importantly, $\tau_j$ and $\tau_j^\dagger$ do not change the $z$-polarization of the $j$'th spin, while $\sigma_j$, $\sigma_j \tau_j$, $\sigma_j^\dagger \tau_j$ and their conjugates change the polarization by one. We will refer to the latter type of operators as {\it shifting} operators.

In contrast to the Ising model, an explicit closed form of $G^n$, $A^n$ is not known \cite{Vernier_2019}. This means that the exact expressions \eqref{solution}, \eqref{solution An} for Heisenberg operators  $G^n_t$, $A^n_t$ can not be immediately converted to  exact expressions for expectation values $\langle G^n\rangle_t$, $\langle A^n\rangle_t$, as in the case of the Ising model. However, it turns out that sums in  eqs. \eqref{solution}, \eqref{solution An}  converge so rapidly that it suffices to calculate a few first  $G^n$, $A^n$ to obtain an excellent approximation to the exact result. Below we demonstrate this by calculating $\langle S_j^z \rangle_t$ for a simple initial pure state $\rho_0=|\Psi_0\rangle\langle \Psi_0 |$ polarized along the $z$-axis,
\be\label{initial condition Potts}
\Psi_0=|\uparrow\uparrow\dots\uparrow\rangle,
\ee
where $|\uparrow\rangle$ is the spin-up state, $S^z|\uparrow\rangle=|\uparrow\rangle$.

We employ computer algebra to calculate the first few $G^n$, $A^n$ starting from $A^0$, $A^1$  according to the recursive relations \eqref{structure of algebra}. The resulting expressions have a considerably more complicated appearance than  those in the Ising case. For example,
\begin{align}\label{G1}
G^1 =& \, \frac49\frac1{1-\omega^*}\sum_{j=1}^N \sigma_j (\tau_j+\tau_j^\dagger-\tau_{j+1}-\tau_{j+1}^\dagger) \sigma_{j+1}^\dagger\,\,-h.c.
\end{align}
The number of essentially different terms necessary to represent higher    $G^n$, $A^n$ rapidly grows: it equals 9 for $A^2$ and $A^{-1}$, 24 for $G^2$, 43 for $A^3$ and $A^{-2}$, 100 for $G^3$,  181 for $A^4$ and $A^{-3}$ , 424 for $G^4$ {\it etc}.  Inspecting these expressions, we conjecture the following properties of $G^n$, $A^n$.
\begin{enumerate}
\item Each  $G^n$, $A^n$ is a sum of {\it strings}, where a string of length $m$ is a tensor product of $m$ operators, each chosen from the set
\be
    \{ \tau, \tau^\dagger, \sigma, \sigma^\dagger, \sigma\tau, \tau^\dagger \sigma^\dagger, \sigma^\dagger \tau, \tau^\dagger \sigma \},
\ee
acting on $m$ consecutive sites.
\item Each string of length $n \geq 2$ has shifting operators  at both its ends.
\item Each $G^n$ consists of strings of lengths not less than $2$.
\item Each $A^n$ with odd $n$ consists of strings of lengths not less than $2$.
\item Each $A^n$ with even $n$ contains strings of length $1$ consisting of operators $\tau_j$, $\tau_j^\dagger$.
\end{enumerate}
We leave the proof (or refutation) of these conjectures for further work. In actual calculations we use only first few   $G^n$, $A^n$ where these  properties have been verified explicitly.

From these properties it follows that $\langle G^n \rangle_0=0$  for all $n$ (in fact, this can be easily derived from the first of the recurrence relations \eqref{generating algebra}) and $\langle A^n \rangle_0=0$ for odd $n$. Further, for even $n$ the only type of string that contributes to $\langle A^n \rangle_0$ is the string of length 1 constructed from $\tau_j$ or $\tau_j^\dagger$. Finally, from the  second line of eq.~\eqref{structure of algebra} with $m=0$, one obtains $\langle A^n \rangle_0=\langle A^{-n} \rangle_0$. As a result, we obtain from eq. \eqref{solution An} a simple expression
\be\label{Sz Potts}
\langle S^z_j \rangle_t=1 - 6\,  a_1^2\,\sum_{l=1}^{\infty}
C^{1(2l-1)}_t  \Big(\langle A^{2l-2}\rangle_0-\langle A^{2l} \rangle_0 \Big)
\ee
with $C^{1(2l-1)}_t$ given by eq. \eqref{C_t}. Note that $a_0$ enters this formula through eqs. \eqref{varepsilon}, \eqref{C_t}. 

In contrast to the Ising case, an explicit general formula for $\langle A^{2l} \rangle_0$ is not known. For this reason we have to truncate the sum in eq. \eqref{Sz Potts} at a finite $l=l_{\rm max}$. Fortunately, the convergence is very rapid. The first few $\langle A^{2l} \rangle_0$ calculating  iteratively from eq. \eqref{generating algebra} with the use of computer algebra read
\renewcommand{\arraystretch}{1.3}
\begin{table}[h!]
\centering
\begin{tabular}{l||c|c|c|c|c}
             $l$ & 0 & $1$ & $2$ & $3 $ & $4 $\\
             \hline
            $N^{-1} \langle A^{2l} \rangle_0$ & $4/3$ & $4/27$ &  $76/729$ & $1636/19683$ & $12452/177147$\\
\end{tabular}
\end{table}

\noindent
We plug these values into eq.\eqref{Sz Potts} and approximately compute $\langle S^z_j \rangle_t$, see Fig. \ref{fig: Potts} (a). We compare four truncations in eq. \eqref{Sz Potts} with $l_{\rm max}=1,2,3,4$ and observe a very rapid convergence, as illustrated in Fig. \ref{fig: Potts} (b).

We conclude this section by emphasizing that a successive application of our method is conditioned to the understanding the Onsager algebra representation for a specific model under consideration. A progress in such understanding in highly desirable.

\section{Summary and outlook \label{sec: discussion}}

To summarize, we have solved coupled Heisenberg equations for a set of operators comprising an Onsager algebra in a system with the  Hamiltonian of the form \eqref{H general}, which itself belongs to this algebra. The solution is given by eqs.~\eqref{solution}--\eqref{C_t}. It allows one to obtain the time evolution of the corresponding observables as soon as the expectation values of these observables in the initial state are known. We have considered two specific realization of the Hamiltonian \eqref{H general}. The first one describes the transverse-field Ising model, where we have calculated  time dependence of an infinite set of observables for a translation-invariant product initial state \eqref{initial condition} with an arbitrary polarisation, see eqs.\eqref{observables} and Fig. \ref{fig}. As a second example, we have studied the superintegrable chiral 3-state Potts model. There we have obtained an approximate but highly accurate description of time evolution of the transverse polarization, see eq. \eqref{Sz Potts} and Fig. \ref{fig: Potts}.

%Several remarks are in order. First, thanks to the fact that the unitary transformation~\eqref{Umn} does not depend on the coefficients $a_0$, %$a_1$ in the Hamiltonian, it can be applied even when these coefficient depend on time. As a result, the problem will be reduced to $(N-1)$ {\it %decoupled} second order linear differential equations with variable coefficients.  This paves the way to a new type of {\it dynamical %integrability} ~\cite{Barmettler_2013,Fioretto_2014,gritsev2017integrable,sinitsyn2018integrable,ermakov2019time,Gamayun_2021_Map}. The latter %can be particularly useful for understanding the many-body Floquet dynamics \cite{gritsev2017integrable, Arze_2020}.

Our approach can be extended to algebras different from the Onsager one. In particular, a set of $\sim 4 N^2$ strictly local operators entering the sums in eq. \eqref{string operators} is closed with respect to commutation. Therefore, one can attempt to derive and solve Heisenberg equations for individual strictly local operators from this set. This will allow one to study the dynamics for non-translation-invariant initial states  and/or transverse field Ising Hamiltonians (in particular, inhomogeneous quantum quench setups \cite{Sotiriadis_2008}) in a site-resolved manner.

An interesting open question is whether the method presented here can be extended to a broader range of integrable model.  Most integrable models are not known to possess a simple algebraic structure analogous to the Onsager algebra (see, however, a recent ref. \cite{miao2021conjectures} where a hidden Onsager algebra has been conjectured for the integrable XXZ spin-$1/2$ chain at the root-of-unity anisotropies).  The absence of such structure prevents a straightforward generalization of the method.

Note, however, that one does not necessarily need an {\it algebra} of operators  to apply the method. Indeed, the requirement of closeness of the set of operators with respect to the commutation is excessive: One actually needs a more weak property of being closed with respect to commutation with the Hamiltonian. Further work will show whether the method can be applied beyond the integrable models with the Onsager algebra.

\section*{Acknowledgements}
I am grateful to O. Gamayun and B. Fine for useful discussions.

% TODO: include author contributions
%\paragraph{Author contributions}
%This is optional. If desired, contributions should be succinctly described in a single short paragraph, using author initials.

% TODO: include funding information
\paragraph{Funding information}
%Authors are required to provide funding information, including relevant agencies and grant numbers with linked author's initials. Correctly-provided data will be linked to funders listed in the \href{https://www.crossref.org/services/funder-registry/}{\sf Fundref registry}.
The work was supported by the Russian Science Foundation under the grant No 17-12-01587.

%\begin{appendix}
%\end{appendix}

% TODO:
% Provide your bibliography here. You have two options:

% FIRST OPTION - write your entries here directly, following the example below, including Author(s), Title, Journal Ref. with year in parentheses at the end, followed by the DOI number.
%\begin{thebibliography}{99}
%\bibitem{1931_Bethe_ZP_71} H. A. Bethe, {\it Zur Theorie der Metalle. i. Eigenwerte und Eigenfunktionen der linearen Atomkette}, Zeit. f{\"u}r Phys. {\bf 71}, 205 (1931), \doi{10.1007\%2FBF01341708}.
%\bibitem{arXiv:1108.2700} P. Ginsparg, {\it It was twenty years ago today... }, \url{http://arxiv.org/abs/1108.2700}.
%\end{thebibliography}

%\bibliographystyle{SciPost_bibstyle} % Include this style file here only if you are not using our template
\bibliography{Ising,C:/D/Work/QM/Bibs/dynamically_integrable,C:/D/Work/QM/Bibs/Floquet}

\nolinenumbers

\end{document}